\documentclass[twocolumn,nofootinbib,aps,preprintnumbers,floatfix]{revtex4}

\usepackage{graphicx}
\usepackage[hypertex]{hyperref}

\newcommand{\nn}{\nonumber}
\newcommand{\beq}{\begin{equation}}
\newcommand{\eeq}{\end{equation}}
\newcommand{\beqa}{\begin{eqnarray}}
\newcommand{\eeqa}{\end{eqnarray}}

\renewcommand{\d}{\mathrm{d}}
\newcommand{\ds}{\displaystyle}
\newcommand{\ov}{\overline}
\newcommand{\GeV}{\ensuremath{\mathrm{GeV}}}
\newcommand{\ord}{\mathcal{O}}

\newcommand{\gsim}{\gtrsim}
\newcommand{\lsim}{\lesssim}
\def\OMIT#1{}%

\newcommand{\ellpm}{\ell^+ \ell^-}
\newcommand{\mbbar}{\overline m_b}
\newcommand{\lqcd}{\Lambda_{\rm QCD}}

\newcommand{\C}{\mathcal{C}}
\newcommand{\Cs}{C_7^\mathrm{incl}{}}
\newcommand{\Cn}{C_9^\mathrm{incl}}

\newcommand{\Csn}{C_{7,9}^\mathrm{incl}}
\newcommand{\Css}{| C_7^\mathrm{incl} |^2}
\newcommand{\Cns}{| C_9^\mathrm{incl} |^2}
\newcommand{\ReCsCn}{\mathrm{Re}\,(C_7^{\mathrm{incl}*}C_9^\mathrm{incl})}

\begin{document}

%%\preprint{\vbox{ \hbox{LBNL-6????} \hbox{hep-ph/yymmnnn} }}

\title{\boldmath Precise predictions for $B\to X_s\ellpm$ in the large
$q^2$ region}

\author{Zoltan Ligeti}

\author{Frank J.\ Tackmann}

\affiliation{Ernest Orlando Lawrence Berkeley National Laboratory,
University of California, Berkeley, CA 94720}

\begin{abstract}

The inclusive $B\to X_s\ellpm$ decay rate in the large $q^2$ region ($q^2 >
m_{\psi'}^2$) receives significant nonperturbative corrections. The resulting
uncertainties can be drastically reduced by normalizing the rate to the $B\to
X_u\ell\bar\nu$ rate with the same $q^2$ cut, which allows for much improved
tests of short distance physics.  We calculate this ratio, including the order
$1/m_b^3$ nonperturbative corrections and the analytically known NNLO
perturbative corrections.  Since in the large $q^2$ region an inclusive
measurement may be feasible via a sum over exclusive states, our results could
be useful for measurements at LHCb and possibly for studies of $B\to X_d\ellpm$.

\end{abstract}

\maketitle

%%%%%%%%%%%%%%%%%%%%%%%%%%%%%%%%%%%%%%%%%%%%%%%%%%%%%%%%%%%%%%%%%%%%%%%%%%%%%%%%
\section{Introduction}
%%%%%%%%%%%%%%%%%%%%%%%%%%%%%%%%%%%%%%%%%%%%%%%%%%%%%%%%%%%%%%%%%%%%%%%%%%%%%%%%

The $b\to s\,\ellpm$ process plays an important role in making overconstraining
measurements of CKM matrix elements and searching for physics beyond the
Standard Model (SM).  This decay has been observed both in inclusive $B\to X_s
\ellpm$~\cite{Aubert:2004it,Iwasaki:2005sy} and exclusive  $B\to K^{(*)}
\ellpm$~\cite{Abe:2004ir,Aubert:2006vb} transitions.  The inclusive $B\to
X_s\ellpm$ decay rate can be calculated in a systematic expansion if one ignores
the $J/\psi$ and $\psi'$ resonances.  It has thus been advocated to compare
calculations and measurements of the (differential) rate for $q^2 <
m_{J/\psi}^2$ and $q^2 > m_{\psi'}^2$, which we shall refer to as the small
$q^2$ and large $q^2$ regions, respectively. Here $q^2 = (p_{\ell^+} +
p_{\ell^-})^2$ is the dilepton invariant mass, and in practice the $q^2$ regions
are chosen as $q^2 \lsim 6\,\GeV^2$ and $q^2 \gsim 14\,\GeV^2$.

The measurements in the two regions are complementary, as they have different
sensitivities to short distance physics, the main theoretical uncertainties have
different origins, and the experimental challenges are also distinct. The most
important operators for $B\to X_s\ellpm$ are
%%%
\beqa
%%O_7 &=& \frac{\alpha_{\rm em}}{2\pi}\, {i q^\nu\over q^2}\,
%%  (\bar s_L \sigma_{\nu\mu} b_R)\, (\bar\ell\, \gamma^\mu \ell)\,, \nn\\
O_7 &=& \frac{e}{16\pi^2}\, \mbbar(\mu)\,
  (\bar s_L \sigma_{\mu\nu} b_R) F^{\mu\nu} , \nn\\
O_9 &=& \frac{\alpha_{\rm em}}{4\pi}\, (\bar s_L \gamma_\mu b_L)\,
  (\bar\ell\, \gamma^\mu \ell)\,, \nn\\
O_{10} &=& \frac{\alpha_{\rm em}}{4\pi}\, (\bar s_L \gamma_\mu b_L)\,
  (\bar\ell\, \gamma^\mu\gamma_5 \ell)\,.
\eeqa
%%%
The operator $O_7$ is important in $B\to X_s\ellpm$ at small $q^2$ due to the
$1/q^2$ pole from the photon propagator (and it dominates the $B\to X_s\gamma$
rate).  At large $q^2$, however, the $O_7$ contribution is small. Compared to
small $q^2$, the rate in the large $q^2$ region has a smaller renormalization
scale dependence and $m_c$ dependence~\cite{Ghinculov:2003qd}. Although the rate
is smaller at large $q^2$, the experimental efficiency is
better~\cite{Aubert:2004it,Iwasaki:2005sy}.  Moreover, requiring large $q^2$
constrains the $X_s$ to have small invariant mass, $m_{X_s}$, which suppresses
the background from $B\to X_c\ell^-\bar\nu \to X_s\ellpm\nu\bar\nu$. To
suppress this background at small $q^2$, an upper cut on $m_{X_s}$ is required,
complicating the theoretical description due to the dependence of the measured
rate on the shape function~\cite{Lee:2005pw}, which is absent at large
$q^2$~\cite{Buchalla:1998mt, Bauer:2000xf}.

Despite these advantages, the large $q^2$ region has been considered less
favored.  The $1/m_b^3$ corrections are not much smaller than the $1/m_b^2$
ones~\cite{Bauer:1999za}, so it is often stated that the $B\to X_s\ellpm$ rate
in the large $q^2$ region has a large hadronic uncertainty~\cite{Bauer:1999za,
Buchalla:1998mt, Hurth:2007xa}. The reason is that the operator product
expansion becomes an expansion in $\lqcd/(m_b-\sqrt{q^2})$~\cite{Neubert:2000ch}
instead of $\lqcd/m_b$.

Our main point is that this uncertainty can be drastically reduced
by comparing measurements and calculations of the ratio
%%%
\beqa\label{ratio}
{\ds\int_{q_0^2}^{m_B^2} {\d\Gamma(B\to X_s \ellpm)\over \d q^2} \over
  \ds\int_{q_0^2}^{m_B^2} {\d\Gamma(B\to X_u\ell\bar\nu)\over \d q^2}}
= \frac{|V_{tb} V_{ts}^*|^2}{|V_{ub}|^2}\,
  \frac{\alpha_{\rm em}^2}{8\pi^2}\, {\cal R}(q_0^2)\,,
\eeqa
%%%
with the same lower cut $q^2 > q_0^2$ in the $b\to s$ and $b\to u$
decays.\footnote{This was noted as a remote possibility in
Ref.~\cite{Bauer:2000xf}, but was subsequently forgotten even by those authors.
The experimental prospects have improved sufficiently that such a study may be
possible in the near future.} The nonperturbative corrections related to the
dominant $O_9$ and $O_{10}$ contributions are the same as for the semileptonic
rate. Thus, as explained below, nonperturbative effects in the ratio in
Eq.~(\ref{ratio}) are suppressed near maximal $q^2$ by
%%%
\beq\label{qual}
1 - \frac{(\C_9 + 2\C_7)^2  + \C_{10}^2}{\C_9^2 + \C_{10}^2} \simeq 0.12\,,
\eeq
%%%
which is nearly an order of magnitude. The scheme we use for the Wilson
coefficients $\C_{7,9,10}$~\cite{Lee:2006gs} will be
defined in Sec.~\ref{sec:results}. Their SM values are
%%%
\beq\label{Civalues}
\C_9 = 4.207\,, \quad \C_{10} = -4.175\,, \quad \C_7 = -0.2611\,.
\eeq
%%%

We calculate in this paper the ratio ${\cal R}(q_0^2)$, which allows one to
translate the measured $B\to X_s\ellpm$ and $B\to X_u\ell\bar\nu$ rates in the
large $q^2$ region to a precision constraint on the Wilson coefficients times
the CKM elements in Eq.~(\ref{ratio}). The normalization of ${\cal R}(q_0^2)$ is
chosen such that at lowest order and in the limit $|\C_7/\C_{9,10}| \ll 1$,
${\cal R}(q_0^2) = \C_9^2 + \C_{10}^2$.  Hereafter we assume the SM and neglect
the strange quark and lepton masses.

%%%%%%%%%%%%%%%%%%%%%%%%%%%%%%%%%%%%%%%%%%%%%%%%%%%%%%%%%%%%%%%%%%%%%%%%%%%%%%%%
\section{\boldmath The $q^2$ spectra to order $1/m_b^3$}
%%%%%%%%%%%%%%%%%%%%%%%%%%%%%%%%%%%%%%%%%%%%%%%%%%%%%%%%%%%%%%%%%%%%%%%%%%%%%%%%

The nonperturbative corrections to the $q^2$ spectrum are calculable in an
operator product expansion (OPE)~\cite{OPE}.  The first corrections appear at
$\ord(\lqcd^2/m_b^2)$~\cite{Falk:1993dh,Ali:1996bm}. They are parameterized by
two nonperturbative matrix elements, $\lambda_1$ and $\lambda_2$. At
$\ord(\lqcd^3/m_b^3)$ there are two new local matrix elements, $\rho_1$ and
$\rho_2$, four time-ordered products, ${\cal T}_{1-4}$~\cite{Mannel:1994kv,
Gremm:1996df}, and process dependent matrix elements of four-quark operators,
$f_i$~\cite{Bigi:1993bh, Blok:1994cd, Voloshin:2001xi}.

The $q^2$ spectrum up to $\ord(\lqcd^3/m_b^3)$ for $B\to X_u\ell\bar\nu$
is given by~\cite{OPE, Gremm:1996df}
%%%
\begin{widetext}
\beqa\label{dGudq2}
\frac{\d\Gamma_u}{\d q^2} &=& \frac{G_F^2 |V_{ub}|^2}{192\, \pi^3}\, m_b^3
  \bigg[ (1-s)^2 (1+2s)\, (2 + \hat\lambda_1)
  + 3(1-15s^2+10s^3)\, (\hat\lambda_2 - \hat\rho_2)
  + \frac{37+24s+33s^2+10s^3}3\, \hat\rho_1 \nn\\*
&&{}\qquad\qquad\quad\ -  {16\over (1-s)_+}\, \hat\rho_1
  - 8\, \delta(1-s)\, (\hat\rho_1 + \hat f_u) \bigg] \,,
\eeqa
%%%
where $s = q^2/m_b^2$, and $1/(1-x)_+ = \lim_{\epsilon\to 0}
[\theta(1-x-\epsilon)/(1-x) + \delta(1-x-\epsilon)\ln \epsilon]$. For $B\to
X_s\ell^+\ell^-$~\cite{Grinstein:1988me, Falk:1993dh, Ali:1996bm,
Bauer:1999za},
%%%
\beqa\label{dGsdq2}
\frac{\d\Gamma_s}{\d q^2} &=& \frac{\Gamma_0}{2}\, m_b^3\,
\bigg\{ (\C_9^2 + \C_{10}^2) \bigg[
  (1-s)^2(1+2s)\, (2 + \hat\lambda_1)
  + 3(1-15s^2+10s^3)\, (\hat\lambda_2 - \hat\rho_2)
  + \frac{37+24s+33s^2+10s^3}3\, \hat\rho_1 \bigg] \nn\\*
&&{}\qquad\quad + 4\, \C_7\, \C_9 \bigg[
  3(1-s)^2\, (2 + \hat\lambda_1)
  - 3(5+6s-7s^2)\, (\hat\lambda_2 - \hat\rho_2)
  + (13+14s-3s^2) \hat\rho_1 \bigg] \nn\\*
&&{}\qquad\quad + {4\, \C_7^2\over s} \bigg[
  (1-s)^2 (2+s)\, (2 + \hat\lambda_1)
  - 3(6+3s-5s^3) (\hat\lambda_2 - \hat\rho_2) +
  {-22+33s+24s^2+5s^3\over 3}\, \hat\rho_1 \bigg] \nn\\*
&&{}\qquad\quad - \big[(\C_9 + 2\C_7)^2 + \C_{10}^2\big]
  \bigg[ {16\over (1-s)_+}\, \hat\rho_1
  + 8\, \delta(1-s)\, (\hat\rho_1 + \hat f_s) \bigg] \bigg\}\,,
\eeqa
\end{widetext}
%%%
where
%%%
\beq\label{Gs}
\Gamma_0 = \frac{G_F^2}{48 \pi^3}\,
  \frac{\alpha_\mathrm{em}^2}{16\pi^2}\,|V_{tb} V_{ts}^*|^2 \,.
\eeq
%%%
The nonperturbative parameters in Eqs.~(\ref{dGudq2}) and (\ref{dGsdq2}) are
%%%
\beqa\label{nonpert}
\hat\lambda_1 &=& \frac{\lambda_1}{m_b^2}
  + \frac{{\cal T}_1 + 3{\cal T}_2}{m_b^3} \,, \qquad
\hat\lambda_2 = \frac{\lambda_2}{m_b^2}
  + \frac{{\cal T}_3 + 3{\cal T}_4}{3m_b^3} \,,\nn\\
\hat\rho_{1,2} &=& \frac{\rho_{1,2}}{m_b^3} \,, \qquad\qquad\qquad\,\,
  \hat f_{u,s} = \frac{f_{u, s}}{m_b^3} \,.
\eeqa
%%%
For our purposes, the ${\cal T}_i$ can be absorbed into $\lambda_{1,2}$. In the
total rate and the $q^2$ spectrum, $\lambda_1$ enters proportional to the $b$
quark decay rate, and the $\rho_2$ contribution is proportional to $\lambda_2$. Hence, the
important nonperturbative parameters for the $q^2$ spectrum are $\lambda_2$,
$\rho_1$, and $f_{u, s}$.

The value of $\lambda_2$ is known fairly precisely, $\lambda_2 =
(m_{B^*}^2-m_B^2)/4 \simeq 0.12\,\GeV^2$. To estimate $\rho_1$, the equations of
motion can be used to relate the relevant operator to a four-quark operator.
Using the vacuum saturation model, $\rho_1 = (2\pi\alpha_s/9) f_B^2 m_B \simeq
(0.4\,\GeV)^3$~\cite{Mannel:1994kv, Bigi:1994ga}. The fits to the $B\to
X_c\ell\bar\nu$ shape variables are sensitive to $\rho_1$ and prefer a larger
central value~\cite{Bauer:2002sh, Abe:2006xq} with significant uncertainties.
We shall use $\rho_1 = (0.1\pm 0.1)\,\GeV^3$.

The four-quark operator contributions, $f_u$ and $f_s$ (sometimes called weak
annihilation, though the light quark flavor need not match the flavor of the
spectator quark), depend on the final state and on the flavor of the decaying
$B$ meson. They contribute near maximal $q^2$, and their contribution has only
been derived for the total rate~\cite{Bigi:1993bh, Blok:1994cd, Bauer:1996ma,
Voloshin:2001xi} and the lepton energy spectrum~\cite{Bigi:1993bh,
Leibovich:2002ys}. However, the four-quark operators have to be consistently
included in the OPE for the fully differential spectrum. This affects the
matching for $\rho_1$ in a nontrivial way at $s=1$, replacing the singular
$\rho_1/(1-s)$ terms present in the earlier literature by the plus distributions
in the last lines of Eqs.~(\ref{dGudq2}) and (\ref{dGsdq2}). Apart from this
unambiguous regularization of the singular integrals at $s=1$, our result in
Eq.~(\ref{dGsdq2}) agrees with Ref.~\cite{Bauer:1999za}.

\begin{figure*}[t]
\centerline{\includegraphics[width=.47\textwidth]{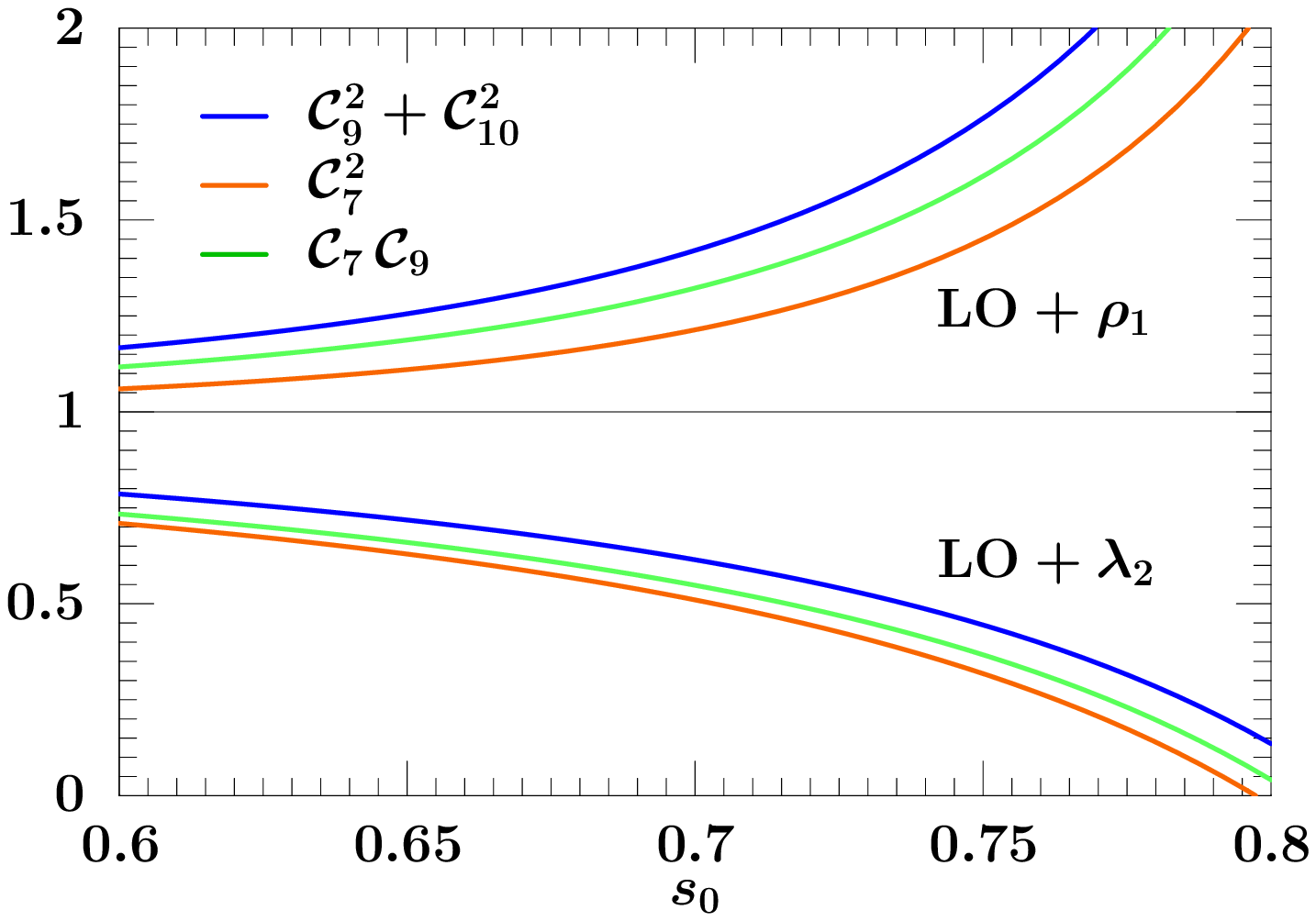}
\hspace{\columnsep}\hfil
\includegraphics[width=.47\textwidth]{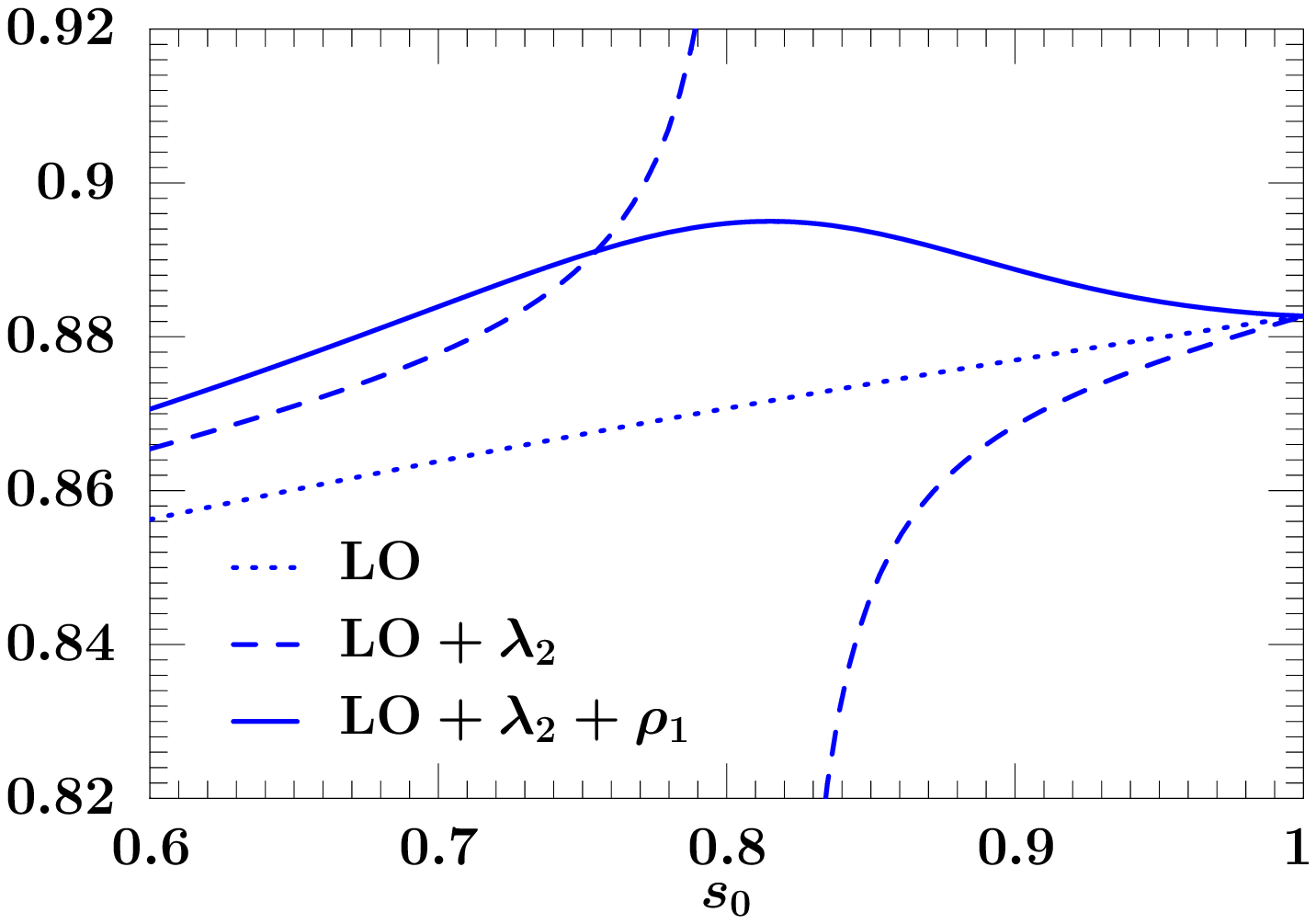}}
\caption{Left: The impact of the corrections proportional to $\lambda_2$
(suppressions) and $\rho_1$ (enhancements) on the $\C_9^2+\C_{10}^2$, $\C_7\,
\C_9$, and $\C_7^2$ contributions to the partonic $B\to X_s\ellpm$ rate
integrated over $s_0 <s< 1$.
Right: The partonic $B\to X_s\ellpm$ rate for $s_0 <s< 1$ divided by the $\C_9^2
+ \C_{10}^2$ contribution (the latter is proportional to the $B\to
X_u\ell\bar\nu$ rate), at lowest order (dotted), including the $\lambda_2$ terms
(dashed), and including both $\lambda_2$ and $\rho_1$ corrections (solid).
(Note the different scales.)}
\label{fig:tree}
\end{figure*}

The values of $f_u$ and $f_s$ are poorly known. They are important, since they
are enhanced by a loop factor, $16\pi^2$. In the notation of
Ref.~\cite{Voloshin:2001xi}, $f_u = 2\pi^2 f_B^2 m_B (B_1 - B_2)$, where
$B_{1,2}$ are phenomenological ``bag parameters''. In the vacuum saturation
model, $B_1 = B_2 = 1$ in charged $B$ decay and $B_1 = B_2 = 0$ in neutral $B$
decay. This gives a significant suppression with large uncertainty, since the
accuracy of the model is poorly known.  Because of this sensitivity to
cancellations between nonperturbative quantities with comparable magnitudes, the
estimates of $f_u$ are uncertain.  On general grounds one expects in charged $B$
decay $f_u^\pm$ to be greater in magnitude than $f_u^0$ in neutral $B$ decay,
since in the former case the spectator flavor matches the flavor of the light
quark in the four-quark operator. The assumption $|B_1-B_2| =
0.1$~\cite{Voloshin:2001xi} for charged $B$ decay leads to $|f_u^\pm| =
0.4\,\GeV^3$. This would give a large uncertainty in ${\cal R}(q_0^2)$,
and we discuss next how it can be reduced.

Flavor $SU(3)$ symmetry implies $f_s^0 \approx f_u^0$. One also expects $f_s^0
\approx f_s^\pm$, though this requires assumptions beyond $SU(3)$ (as there are
two singlets in $3\times \ov 3 \times 3 \times \ov 3$). Thus, one may use in
Eq.~(\ref{ratio}) the average of the charged and neutral $B\to X_s\ellpm$ rate
and the neutral $B^0\to X_u\ell\bar\nu$ rate in the presence of the same $q^2$
cuts.\footnote{For $B\to X_d\ellpm$, $SU(3)$ implies $f_d^0 \approx f_u^\pm$ and
$f_d^\pm \approx f_u^0$.  If semileptonic $B_s\to X_u\ell\bar\nu$ decays can be
measured in the large $q^2$ region, that would open up yet another set of
applications.}  Separately measuring $B^0\to X_u\ell\bar\nu$ and $B^\pm\to
X_u\ell\bar\nu$ in the large $q^2$ region is important, even without the
additional reasons discussed here, for the determination of $|V_{ub}|$ and to
constrain weak annihilation.  We anticipate that this separation will be available
by the time the $B\to X_s\ellpm$ rate in the large $q^2$ region is precisely
measured.

To illustrate the slow convergence of the OPE at large $q^2$, we show in the
left plot in Fig.~\ref{fig:tree} the effect of the dominant $1/m_b^2$ and
$1/m_b^3$ corrections, proportional to $\lambda_2$ and $\rho_1$, respectively.
For each of the $\C_9^2+\C_{10}^2$, $\C_7\, \C_9$, and $\C_7^2$ contributions,
we plot the free quark decay rate plus either the $\lambda_2$ (curves below
unity) or the $\rho_1$ (curves above unity) terms integrated over $s_0 < s < 1$,
normalized to the free quark decay rate. The $\lambda_2$ term, which is about a
$-2\%$ correction to the total rate, makes the spectrum negative for $s \gsim
0.9$, and the integrated rate negative for $s_0 \gsim 0.8$.  The terms
proportional to $\rho_1$ give a correction of comparable magnitude and opposite
sign (taking $\rho_1 = 0.1\,\GeV^3$). Even at the rather low cut, $s_0 = 0.6$,
the correction to the rate proportional to $\lambda_2$ is about $-21\%$, while
that proportional to $\rho_1$ is about $+17\%$. Moreover, these estimates do not
include the $f_i$ terms discussed above.  Thus, even for the rate integrated
over the entire large $q^2$ region, it is not clear how to assign a robust
uncertainty.

The important point is that while these corrections are large, the terms
proportional to $\C_9^2 + \C_{10}^2$ dominate at each order, and they are
identical to those that occur in $B\to X_u\ell\bar\nu$.  We present a detailed
numerical study in the next section.  To illustrate the point, the right plot in
Fig.~\ref{fig:tree} shows the partonic $B\to X_s\ellpm$ rate integrated over
$s_0 <s< 1$ divided by the $\C_9^2+\C_{10}^2$ contribution, at lowest order
(dotted), including the $\lambda_2$ terms (dashed), and including both
$\lambda_2$ and $\rho_1$ corrections (solid). The blowup of the dashed curve
near $s_0 \sim 0.8$ is unphysical and occurs because the rate truncated at
$\ord(\lqcd^2/m_b^2)$ becomes negative at slightly different values of $s_0$ for
the $\C_9^2+\C_{10}^2$, $\C_7\, \C_9$, and $\C_7^2$ contributions, as can be
seen in the left plot. As expected, for $s_0 \to 1$ the ratio is $0.88$ at each
order, i.e., unity minus the $0.12$ correction in Eq.~(\ref{qual}).

%%%%%%%%%%%%%%%%%%%%%%%%%%%%%%%%%%%%%%%%%%%%%%%%%%%%%%%%%%%%%%%%%%%%%%%%%%%%%%%%
\section{\boldmath Results for ${\cal R}(q_0^2)$}
\label{sec:results}
%%%%%%%%%%%%%%%%%%%%%%%%%%%%%%%%%%%%%%%%%%%%%%%%%%%%%%%%%%%%%%%%%%%%%%%%%%%%%%%%

To organize the different short-distance contributions,
and combine the nonperturbative and the next-to-next-to-leading order (NNLO)
perturbative corrections, we use the scheme introduced in Ref.~\cite{Lee:2006gs}.
The Wilson coefficients $\C_{7,9,10}$ are defined as
%%%
\beqa\label{Ci_intro}
\C_7 &=& C_7(\mu)\, \big[\mbbar(\mu)/m_b\big] + \ldots \,,\nn\\
\C_9 &=& C_9(\mu) + \ldots \,,\nn\\
\C_{10} &=& C_{10} \,.
\eeqa
%%%
where the ellipses denote a minimal set of perturbative corrections, such that
$\C_{7,9}$ are $\mu$ independent and real in the SM (which is automatic for
$C_{10}$). We use the $1S$ scheme~\cite{Hoang:1998ng}, which improves the
behavior of the perturbative expansions (for the semileptonic $q^2$ spectrum
both the $\alpha_s^2\beta_0$~\cite{Bauer:2000xf} and full
$\alpha_s^2$~\cite{Czarnecki:2001cz} corrections are known). Consequently, we
use $m_b \equiv m_b^{1S}$ everywhere, except for the $\ov{\rm MS}$ $b$-quark
mass, $\mbbar(\mu)$, in Eq.~(\ref{Ci_intro}), which is renormalized together
with $C_7(\mu)$.

The inclusive decay rate is expressed in terms of the effective Wilson
coefficients~\cite{Lee:2006gs}
%%%
\beqa\label{Cincl}
\Cs(q^2) &=& \C_7 + F_7(q^2) + G_7(q^2) \,,\nn\\*
\Cn(q^2) &=& \C_9 + F_9(q^2) + G_9(q^2) \,,
\eeqa
%%%
which are defined such that all terms on the right-hand side are separately
$\mu$ independent to the order we are working at. We view the coefficients
$\C_{7,9,10}$ as the parameters sensitive to physics beyond the SM, which should
be extracted from experimental data and compared with their SM predictions. Even
if they receive significant new physics contributions, the functions
$F_{7,9}(q^2)$ and $G_{7,9}(q^2)$ are likely to be dominated by the SM.

The $F_{7,9}(q^2)$ terms in Eq.~(\ref{Cincl}) contain contributions from the
remaining $O_{1-6,8}$ operators in the effective Hamiltonian, for which we
employ a partial NNLO treatment. We use the Wilson coefficients at
$\ord(\alpha_s)$~\cite{Adel:1993ah, Chetyrkin:1996vx, Bobeth:1999mk,
Bobeth:2003at, Gambino:2003zm}, but only keep the $\ord(\alpha_s^0)$
contributions to $F_9(q^2)$~\cite{Grinstein:1988me, Misiak:1992bc}. (Note that
$F_7(q^2)$ vanishes at order $\alpha_s^0$.) We cannot include the
$\ord(\alpha_s)$ corrections to $F_{7,9}(q^2)$, because the dominant $O_{1,2}$
contributions are only known analytically in the small $q^2$
region~\cite{Asatryan:2001de}. They have been evaluated numerically for large
$q^2$, and lead to a reduced scale dependence and central
value~\cite{Ghinculov:2003qd}.

The $G_{7,9}(q^2)$ terms in Eq.~(\ref{Cincl}) contain the calculable
$\lqcd^2/m_c^2$ nonperturbative corrections associated with intermediate $c\bar
c$ loops~\cite{Voloshin:1996gw, Buchalla:1997ky}. They can be included this way
for any differential rate~\cite{Lee:2006gs}, provided $\ord(\alpha_s/m_c^2,\,
1/m_c^4)$ cross terms are neglected. They affect ${\cal R}(q_0^2)$ slightly
below the $1\%$ level in the large $q_0^2$ region.

Thus, the $q^2$ spectra, including up to NNLO perturbative and $1/m_b^3$
nonperturbative corrections, are
%%%
\beqa\label{q2rates}
\frac{\d\Gamma_u}{\d q^2} &=& \frac{G_F^2 |V_{ub}|^2}{96\, \pi^3}\, m_b^3
  (1 - s)^2 \big[1 + 2s - \Omega^{99}(s)\big]
+ \frac{\d\Gamma_u^{1/m}}{\d q^2} ,\nn\\*
%%%
\frac{\d\Gamma_s}{\d q^2} &=& \Gamma_0 m_b^3\, (1 - s)^2
  \bigg\{ \big(\Cns+\C_{10}^2\big) \big[1 + 2s - \Omega^{99}(s)\big] \nn\\*
&&{} + 4\,\ReCsCn\, \big[3 - \Omega^{79}(s)\big] \\*
&&{} + \frac{4 \Css}{s} \big[2 + s - \Omega^{77}(s)\big]
  + \Gamma^\mathrm{brems} \! \bigg\}
  + \frac{\d\Gamma_s^{1/m}}{\d q^2} . \nn
\eeqa
%%%
The power suppressed corrections, $\d\Gamma_{u,s}^{1/m}/\d q^2$, are given by
the terms in Eqs.~(\ref{dGudq2}) and (\ref{dGsdq2}) proportional to
$\lambda_{1,2}$, $\rho_{1,2}$, and $f_{u,s}$, appropriately replacing $\C_{7,9}$
by $\Csn$ as in Eq.~(\ref{q2rates}). The functions $\Omega^{ij}(s)$,
%%%
\beqa
\Omega^{99}(s) &=& \frac{\alpha_s C_F}{2\pi}\,
  \big[\omega_L^{99}(s) + 2s\,\omega_T^{99}(s)\big]
,\nn\\*
\Omega^{77}(s) &=& \frac{\alpha_s C_F}{2\pi}\,
  \big[s\,\omega_L^{77}(s) + 2\,\omega_T^{77}(s)\big]
,\nn\\*
\Omega^{79}(s) &=& \frac{\alpha_s C_F}{2\pi}\,
  \big[\omega_L^{79}(s) + 2\,\omega_T^{79}(s)\big] ,
\eeqa
%%%
contain the $\ord(\alpha_s)$ corrections to the matrix elements of the $O_i O_j$
contribution~\cite{Grinstein:1988me, Misiak:1992bc, Asatryan:2001de,
Ghinculov:2002pe, Asatrian:2002va} converted to the $1S$ scheme, with
$\omega^{ij}_{L,T}(s)$ given in Ref.~\cite{Lee:2006gs}. We neglect finite
bremsstrahlung corrections, $\Gamma^\mathrm{brems}$, associated with
$O_{1-6,8}$, because they are negligible at large $q^2$~\cite{Asatryan:2002iy,
Asatrian:2003yk}.

The perturbative uncertainty due to the choice of renormalization scale for
$\alpha_{\rm em}$, which appears in the prefactor of $\mathcal{R}(q_0^2)$
in Eq.~(\ref{ratio}), can be eliminated by including the relevant higher order
electroweak corrections to the $B\to X_s\ellpm$ rate, which have been studied in
Refs.~\cite{Bobeth:2003at, Huber:2005ig} in the small $q^2$ region.

\begin{table}[t]
\tabcolsep 6pt
\begin{tabular}{c|cc}
\hline\hline
parameter  &  central value  &  uncertainty \\
\hline\hline
$\mu\ [\GeV]$  &  4.7  &  ${}^{+4.7}_{-2.35}$ \\
$m_b\ [\GeV]$  &  4.7  &  $\pm 0.04$ \\
$m_c\ [\GeV]$  &  1.41  & $\pm 0.05$ \\
\hline
$\lambda_2\ [\GeV^2]$  &  0.12  & $\pm 0.02$ \\
$\rho_1\ [\GeV^3]$  &  0.1  &  $\pm 0.1$ \\
$f_u^\pm\ [\GeV^3]$  &  0  &  $\pm 0.4$ \\
$f_u^0 - f_s\ [\GeV^3]$  &  0  &  $\pm 0.04$ \\
$f_u^0 + f_s\ [\GeV^3]$  &  0  &  $\pm 0.2$ \\
\hline\hline
\end{tabular}
\caption{Central values and ranges of input parameters. The parameters
$\lambda_1$, $\rho_2$, $\mathcal{T}_i$ are irrelevant for this work and are set
to their central values, $\lambda_1 = -0.27\,\GeV^2$ and $\rho_2 = \mathcal{T}_i
= 0$.}
\label{tab:inputs}
\end{table}

We present our result for the SM prediction for two different values of $q_0^2$,
as it is an open question what its most suitable choice is. Using the input
values in Table~\ref{tab:inputs}, with all other inputs as in
Ref.~\cite{Lee:2006gs}, we obtain
%%%
\beqa\label{RCis}
{\cal R}(14\,\GeV^2) &=& \C_9^2 + \C_{10}^2 + 4.79\,\C_7^2 + 4.31\,\C_7\,\C_9
\nn\\
&&{} + 1.06\,\C_9 + 2.24\, \C_7 + 0.95\,, \nn\\*
{\cal R}(15\,\GeV^2) &=& \C_9^2 + \C_{10}^2 + 4.27\,\C_7^2 + 4.10\,\C_7\,\C_9
\nn\\
&&{} + 0.97\,\C_9 + 1.91\,\C_7 + 0.93\,.
\eeqa
%%%
Using the central values of the $\C_i$ in Eq.~(\ref{Civalues}) and evaluating
the uncertainties by varying the parameters within their ranges given in
Table~\ref{tab:inputs}, we find
%%%
\beqa\label{Rvalues}
{\cal R}(14\,\GeV^2) = 35.55\, \big(
1 &\pm& 0.046_{[f_{u,s}]} \pm 0.012_{[\lambda_2,\rho_1]}
  \nn\\*
&\pm& 0.054_{[\mu]} \pm 0.030_{[\C_i]} \big) \,, \nn\\*
{\cal R}(15\,\GeV^2) = 35.42\, \big(
1 &\pm& 0.065_{[f_{u,s}]} \pm 0.016_{[\lambda_2,\rho_1]}
  \nn\\*
&\pm& 0.051_{[\mu]} \pm 0.030_{[\C_i]} \big) \,.
\eeqa
%%%
Eqs.~(\ref{RCis}) and (\ref{Rvalues}) can be directly compared with experimental
measurements to constrain the Wilson coefficients (mainly $\C_9^2 + \C_{10}^2$)
and test the standard model.

\begin{figure}[t]
\includegraphics[width=.95\columnwidth]{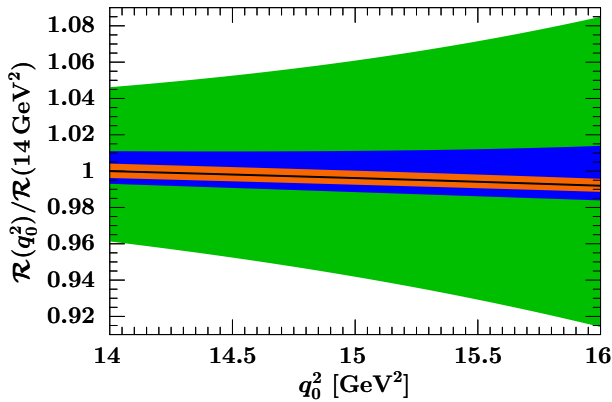}
\caption{Nonperturbative uncertainties in ${\cal R}(q_0^2)/{\cal
R}(14\,\GeV^2)$. The black curve is the central value. The green (wide light),
blue (dark), and orange (narrow light) shaded bands show the uncertainties from
$f_{u,s}$, $\rho_1$, and $\lambda_2$, respectively.}
\label{fig:2}
\end{figure}

The first two uncertainties in Eqs.~(\ref{Rvalues}) are due to nonperturbative
corrections.  They are shown in Fig.~\ref{fig:2}, where we plot ${\cal R}(q_0^2)
/ {\cal R}(14\,\GeV^2)$ as a function of $q_0^2$. The most important uncertainty
is due to the four-quark operators (weak annihilation), in particular, the
difference $f_u - f_s$, which does not cancel in the ratio ${\cal R}(q_0^2)$.
The green (wide light) region in Fig.~\ref{fig:2} corresponds to allowing
$|f_u^0 - f_s| < 0.04\,\GeV^3$ and $|f_u^0 + f_s| < 0.2\,\GeV^3$, which is
appropriate if the average of the $B^0$ and $B^\pm$ rare decay data is compared
with the $B^0$ semileptonic data. Fixing $f_u^0 = f_s^0$, which holds in the
$SU(3)$ limit, and varying them together in the range $|f_{u,s}^0| <
0.1\,\GeV^3$, gives a very small uncertainty, below $1\%$. If $B^0\to
X_u\ell\bar\nu$ is not measured separately from $B^\pm\to X_u\ell\bar\nu$, the
uncertainty in ${\cal R}(14\,\GeV^2)$ increases to above $20\%$ (taking $|f_u| <
0.2\,\GeV^3$ and $f_s=0$ as an estimate for the average of the charged and
neutral modes).

Even with our conservative range for $\rho_1$, the combined uncertainty from
$\rho_1$ and $\lambda_2$ is very small, at the 1\% level. The allowed variation
of $\lambda_2$ accounts for ${\cal T}_{3,4}$, which it absorbs in
Eq.~(\ref{nonpert}), and for $\rho_2$, which it is proportional to in
Eqs.~(\ref{dGudq2}) and (\ref{dGsdq2}). The individual uncertainties from
$\rho_1$ and $\lambda_2$ are shown in Fig.~\ref{fig:2} as the blue (dark) and
orange (narrow light) bands, respectively. The total uncertainty from
nonperturbative corrections, estimated conservatively, is only about $6\%$.
Without normalizing to the $B\to X_u\ell\nu$ rate with the same cut, the
uncertainty in the $B\to X_s\ellpm$ rate for $q^2 \geq 14\,\GeV^2$ due to
$\lambda_2$, $f_s$, and $\rho_1$ would be $4\%$, $9\%$, and $21\%$,
respectively.

\begin{figure}[t]
\includegraphics[width=.95\columnwidth]{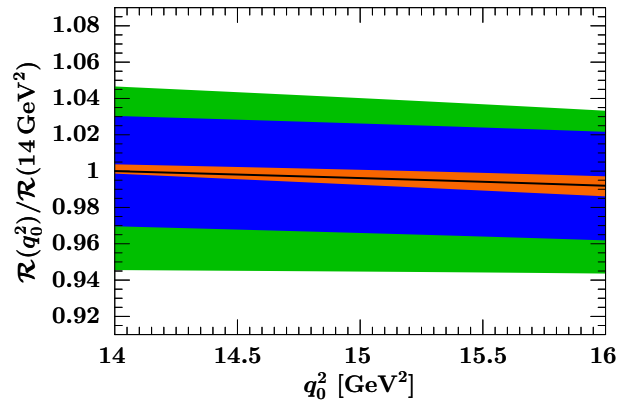}
\caption{Perturbative uncertainties in ${\cal R}(q_0^2)/{\cal R}(14\,\GeV^2)$.
The black curve is the central value (as in Fig.~\ref{fig:2}). The orange
(narrow light) and green (wide light) bands show the $\mu$ dependence (see
text). The blue (dark) band is the uncertainty from $\C_{7,9,10}$.}
\label{fig:3}
\end{figure}

The remaining two uncertainties in Eqs.~(\ref{Rvalues}) are due to the
perturbative corrections, and are illustrated in Fig.~\ref{fig:3}. The variation
in $\mu$, shown by the green (wide light) band, is almost entirely due to the
$\ord(\alpha_s^0)$ pieces in $F_9(q^2)$, because their $\mu$ dependence does not
cancel in $\mathcal{R}(q_0^2)$. We expect that it will be reduced to below $1\%$
by including the full NNLO result. This is shown by the narrow orange (narrow
light) region in Fig.~\ref{fig:3}, obtained by including the subset of
$\alpha_s$ corrections to $F_9(q^2)$ which cancels its leading $\mu$ dependence
(see Eq.~(A5) in Ref.~\cite{Lee:2006gs}). Finally, the blue (dark) band shows
the uncertainty from $\C_{7,9,10}$, which includes their residual $\mu$
dependence and dependence on the electroweak matching scale (mainly affecting
$\C_9$), as well as their dependence on the top-quark mass (mainly affecting
$\C_{10}$). The uncertainty in $F_{7,9}(q^2)$ due to the top quark mass is
negligible and that due to the electroweak matching scale is well below $1\%$.

The uncertainties due to other input parameters are much smaller than those
shown in Eq.~(\ref{Rvalues}). The uncertainty from $m_c$ is well below $1\%$,
about the size of that of $\lambda_2$, shown by the orange (narrow light) band
in Fig.~\ref{fig:2}. The $m_b$ dependence is negligible, because it almost
completely cancels between the numerator and denominator of ${\cal R}(q_0^2)$.
This is another significant advantage of considering the ratio
$\mathcal{R}(q_0^2)$. Both integrated rates with a cut at $q_0^2 = 14\,\GeV^2$
scale roughly as $m_b^{13}$, yielding a $11\%$ uncertainty. Normalizing the
$B\to X_s\ellpm$ rate to the total $B\to X_u\ell\bar\nu$ rate (proportional to
$m_b^5$), would still leave about an $m_b^8$ dependence (and even stronger if
normalized to $B\to X_c\ell\bar\nu$), which would give an additional $7\%$
uncertainty compared to our results.

%%%%%%%%%%%%%%%%%%%%%%%%%%%%%%%%%%%%%%%%%%%%%%%%%%%%%%%%%%%%%%%%%%%%%%%%%%%%%%%%
\section{Concluding remarks}
\label{sec:conclusions}
%%%%%%%%%%%%%%%%%%%%%%%%%%%%%%%%%%%%%%%%%%%%%%%%%%%%%%%%%%%%%%%%%%%%%%%%%%%%%%%%

In this paper we pointed out that the theoretical uncertainty of the $B\to
X_s\ellpm$ rate in the large $q^2$ region, which is dominated by nonperturbative
uncertainties, can be significantly reduced by normalizing the $B\to X_s\ellpm$
rate to the $B\to X_u\ell\bar\nu$ rate with the same $q^2$ cut. Fully exploiting
this proposal requires the experimental separation of $B^0\to X_u\ell\bar\nu$
and $B^\pm\to X_u\ell\bar\nu$ in the large $q^2$ region to eliminate the largest
part of the four-quark operator (weak annihilation) contributions.

In much of the theory literature $B\to X_s \ellpm$ has been normalized to $B\to
X_c\ell\bar\nu$ and sometimes to $B\to X_u\ell\bar\nu$. However, to achieve the
most reduction of theoretical uncertainties, related to both the matrix elements
of the higher dimension operators in the OPE and the value of $m_b$, the
normalization should be done using the $B\to X_u\ell\bar\nu$ rate with the same
cuts as in $B\to X_s \ellpm$.  This holds both in the small $q^2$
region~\cite{Lee:2005pw}, and especially in the large $q^2$ region studied in
this paper.

An uncertainty we have not addressed explicitly is due to higher $c\bar c$
resonances, $\psi(3S)$ and above, but their contributions are much smaller than
those of the $\psi$ and $\psi'$ and are not expected to introduce a significant
uncertainty.  Improvements in both theory and experiment will determine what is
the optimal choice of the lower limit, $q_0^2$, of the large $q^2$ region to
minimize the total uncertainty. If $q_0^2$ is increased much above
$m_{\psi'}^2$, one should get concerned about quark-hadron duality. Even for
$q_0^2 = 14\,\GeV^2$ the range of hadronic invariant masses summed over is only
$m_{X_s} \leq 1.53\,\GeV$ (and $m_{X_s} \leq 1.41\,\GeV$ for $q_0^2 =
15\,\GeV^2$). Hadronic $\tau$ decay data~\cite{Davier:2005xq} indicate that
duality may already be a good approximation at these values.

In the small $q^2$ region, in addition to the $q^2$ spectrum and the
forward-backward asymmetry, $\d A_{\rm FB}/\d q^2$, one can constrain a third
linear combination of the Wilson coefficients by splitting the rate into
``transverse" and ``longitudinal" components~\cite{Lee:2006gs}. At large $q^2$
this is not advantageous, since both are dominated by the $\C_9^2 + \C_{10}^2$
contribution, and thus yield very similar constraints. Our results can also be
applied to the forward-backward asymmetry, $A_{\rm FB}$, which at large $q^2$ is
mainly sensitive to $\C_9\,\C_{10}$. As noted in Ref.~\cite{Bauer:1999za}, the
OPE at large $q^2$ appears to be significantly better behaved for $A_{\rm FB}$
than for the $q^2$ spectrum.  However, if only the normalized forward-backward
asymmetry, $(\d A_{\rm FB}/\d q^2) / (\d\Gamma/\d q^2)$, is measured
in the large $q^2$ region (which may be the least difficult experimentally), it
would inherit the large uncertainty of the rate.  In this case, one could
normalize this measurement to the rate or the normalized $A_{\rm FB}$ in $B\to
X_u\ell\bar\nu$ with the same $q^2$ cut. The latter should be accessible with the
large samples of fully reconstructed $B$ decays used to extract $|V_{ub}|$.

It is not yet known if an inclusive study of $B\to X_d\ellpm$ can ever be
carried out, but it may be less difficult in the large $q^2$ than in the small
$q^2$ region.  At large $q^2$ a semi-inclusive experimental analysis might
become feasible at a super $B$ factory or even at LHCb. The methods discussed in
this paper are clearly applicable to this decay as well. Moreover, in the large
$q^2$ region the exclusive rates can be understood model independently using
continuum methods~\cite{Grinstein:2004vb,Ligeti:1995yz} or lattice QCD.

To summarize, we showed that it is possible to gain theoretically clean short
distance information from the large $q^2$ region of $B\to X_s\ellpm$. For this,
the experimentally most important input, in addition to precisely measuring
$B\to X_s\ellpm$ at large $q^2$, is the measurement of $B^0\to X_u\ell\bar\nu$
with the same $q^2$ cut without averaging with $B^\pm\to X_u\ell\bar\nu$.  On
the theory side, it is desirable to include the full NNLO calculation,
$F_{7,9}(q^2)$ at $\ord(\alpha_s)$, which will largely reduce the perturbative
uncertainties in Fig.~\ref{fig:3}, and leave us with a prediction for ${\cal
R}(q_0^2)$ in the 14\,--15\,\GeV\ region with a theoretical uncertainty about
the 5\% level.

\acknowledgments

We thank Christian Bauer, Iain Stewart, and Francesca Di Lodovico for helpful
conversations.  We thank Tobias Hurth for pointing out an error in
Eq.~(\ref{dGsdq2}) in an earlier version of this paper.
This work was supported in part by the Director, Office of
Science, Office of High Energy Physics of the U.S.\ Department of Energy under
the Contracts DE-AC02-05CH11231.

\end{document}